# Accreting Binary Eccentricities Follow Predicted Equilibrium Values

Allen R. Murray and Paul C. Duffell
Department of Physics and Astronomy, Purdue University, 525 Northwestern Avenue, West Lafayette, IN 47907, USA; murray92@purdue.edu, pduffell@purdue.edu

## Abstract

We investigate observations of circumbinary disks (CBDs) to find evidence for an equilibrium eccentricity predicted by current binary accretion theory. Although stellar binary demographics in the Milky Way show no evidence for a preferred eccentricity for binary systems, we show that actively accreting systems lie on a predicted equilibrium eccentricity curve. We constrain our sample to only systems that have well-defined orbital parameters (e.g., eccentricity, mass ratio, inclination angle). We find observations are consistent with theory for stellar binaries that are aligned with the disk and that are separated enough that tidal circularization is negligible. This suggests that eccentricity in these systems evolves after the dissipation of the CBD, given the flat eccentricity distribution of binary systems in the Milky Way.

*Unified Astronomy Thesaurus concepts:* Binary stars (154); Stellar accretion disks (1579); Stellar accretion (1578); Protoplanetary disks (1300)

## 1. Introduction

Circumbinary disks (CBDs) are a natural phenomenon occurring in many astrophysical contexts. They are thought to form after galaxy mergers drive accretion onto supermassive black holes binaries (SMBHBs; e.g., M. C. Begelman et al. 1980; M. Milosavljević & D. Merritt 2001; A. Escala et al. 2005; M. Milosavljević & E. S. Phinney 2005; M. Dotti et al. 2007; L. Mayer et al. 2007; J. Cuadra et al. 2009; D. Chapon et al. 2013), through fallback accretion around a massive giant star and white dwarf companion (e.g., A. Kashi & N. Soker 2011; S. Tuna & B. D. Metzger 2023), and through disk fragmentation in young binary star evolution (e.g., A. P. Boss 1986; I. A. Bonnell & M. R. Bate 1994a, 1994b; K. M. Kratter et al. 2008). The latter will be the primary focus of this paper.

Current theories of disk–binary interactions predict that a binary system with a CBD will see its mass ratio, eccentricity, and semimajor axis evolve over time as the disk accretes (for mass ratio considerations, see P. C. Duffell et al. 2020 and D. J. Muñoz et al. 2020; for eccentricity and semimajor axis analysis, see M. Siwek et al. 2023). It was shown (M. S. L. Moody et al. 2019; D. J. Muñoz et al. 2019) that though accretion and gravitational torques are of the same magnitude for order unity mass ratio systems, in many systems, accretion drives a net positive torque on the binary. This is due to the larger torque that the circumstellar minidisks exert on the individual stars. This effect causes the binary to drift apart rather than together, for many binary–disk setups. (e.g., Y. Tang et al. 2017; M. S. L. Moody et al. 2019; D. J. Muñoz et al. 2019, 2020). Accretion is also predicted to evolve the binary mass ratio, (M. R. Bate & I. A. Bonnell 1997; B. D. Farris et al. 2014), which is thought to give rise to observed "twin" binaries ($q \gtrsim 0.95$; K. El-Badry et al. 2019).

Eccentricity evolution has been explored by other studies (D. J. D'Orazio & P. C. Duffell 2021; J. Zrake et al. 2021) finding an attractor eccentricity of $e \approx 0.4$–$0.5$ for CBD systems with a wide range of mass ratios and other parameters.

J. Zrake et al. (2021) also found preferential eccentricities for post–asymptotic giant branch binaries of $e < 0.05$ and $e \approx 0.3$.

The predictions of an equilibrium or "attractor" eccentricity were expanded by the study of M. Siwek et al. (2023). M. Siwek et al. (2023) expanded the parameter space to include a range of binary systems with varying mass ratio $q$ and eccentricity $e_b$, making predictions for the time derivative of $e_b$ as a function of both parameters. They found that this "'attractor" eccentricity is a function of the mass ratio of the binary, providing an equilibrium curve that all binaries should be attracted to.

Much of the above theoretical work was developed to predict the evolution of SMBHB populations. In this study, we apply this theory to actively accreting stellar binary systems and interpret the eccentricity distributions of observed binary systems. Theoretical studies have predicted stable eccentricities for a variety of systems, which raises the question: Why do we not see any of these trends in the data from population surveys such as GAIA (K. El-Badry et al. 2019) and APOGEE (A. M. Price-Whelan et al. 2020)? Typical binary populations (e.g., Figure 7 of A. M. Price-Whelan et al. 2020), show a pileup of systems with $e \approx 0$ and short orbital period ($P < 10$ days) that is indicative of tidal circularization. Otherwise, these populations show no preferential eccentricity for binary systems and a flat distribution across $0.1 \lesssim e \lesssim 0.7$.

In this study, rather than look at general binary star systems in the Milky Way, we investigate a population of actively accreting binary systems, using the catalog of I. Czekala et al. (2019). Our goal is to investigate the alignment and eccentricity distribution of these systems to search for evidence of binary–disk evolution predicted by theory.

The paper is structured as follows: Section 2 will provide our data categorization process, Section 3 will describe the theory we used to motivate the project, Section 4 will detail our results and their interpretation, and Section 5 will give conclusions and potential follow-up studies.

## 2. Data Categorization

Using Tables 3, 4, and 5 from I. Czekala et al. (2019), we selected systems with well-defined eccentricity, $e$; mass ratio,

**Table 1**
Observed Binary–Disk Systems Used in This Study with Well-defined Binary Eccentricity, $e$; Mass Ratio, $q$; Binary–Disk Inclination, $i$ (in Degrees); and Inclination Uncertainty, $\Delta i$

| System Name | $e$ | $q$ | $i$ | $\pm \Delta i$ |
|---|---|---|---|---|
| Tidally Circularized | | | | |
| V4046 Sgr | 0.0 | 0.94 | 0.0 | 2.3 |
| CoRoT 2239 | 0.0 | 0.74 | 0.0 | 5.0 |
| Planar Aligned | | | | |
| HD 131511 | 0.51 | 0.57 | 0.0 | 15 |
| $\alpha$CrB | 0.37 | 0.36 | 0.0 | 35 |
| TWA 3A | 0.628 | 0.84 | 0.0 | 25 |
| HD 200775 | 0.30 | 0.82 | 0.0 | 20 |
| Ak Sco | 0.47 | 1.00 | 0.0 | 2.7 |
| DQ Tau | 0.57 | 0.94 | 0.0 | 2.7 |
| UZ Tau E | 0.33 | 0.29 | 0.0 | 2.7 |
| Hierarchical Trinary | | | | |
| HD 142527 | 0.50 | 0.05 | 35 | 5.0 |
| GG Tau Aa-Ab | 0.50 | 0.89 | 25 | 5.0 |
| GW Ori A-B | 0.13 | 0.60 | 50 | 5.0 |
| GW Ori AB-C | 0.13 | 0.26 | 45 | 5.0 |
| Polar Aligned | | | | |
| 99 Her | 0.77 | 0.49 | 80 | 6.0 |
| HD 98800B | 0.78 | 0.86 | 92 | 3.0 |

**Note.** Data from I. Czekala et al. (2019).

$q$; and mutual inclination, $\theta$. We will use $i$ for inclination going forward. Our complete set of binary–disk systems is shown in Table 1. We left out V892 Tau due to the ambiguity in orbit orientation.

The data set in Table 1 is sectioned into four categories. For systems in the source's tables without traditional error bars (i.e., $i < 10°$), we set the inclination as $i = 0$ and took the upper limit as the uncertainty in $i$. These systems we refer to as "aligned." Systems with approximately 90° inclination are identified as "polar aligned." Binaries on short orbits with $e = 0$ are expected to be circularized by tides. Hierarchical trinaries are the systems with a third body.

### 3. Theoretical Interpretation of Data

This section will detail the theory we used to interpret the observational data. Due to the turbulent nature of the molecular clouds that star systems are commonly formed in, CBD could form misaligned from the binary orbital plane. (e.g., J. L. Monin et al. 2007; M. R. Bate et al. 2010). Investigating the parameters needed to evolve CBDs to polar alignment, R. G. Martin & S. H. Lubow (2017) used three-dimensional hydrodynamical models. They found that, for equal mass binaries, the disk will evolve to polar alignment given certain initial binary eccentricity $e$ and mutual binary–disk inclination $i$. For further discussion about conditions for polar alignment, see R. G. Martin & S. H. Lubow (2018). Equation (33) from R. G. Martin & S. H. Lubow (2019) provides a criterion for whether a disk will evolve toward polar or planar alignment. To find the absolute minimum mutual inclination angle needed for polar alignment, we assumed a disk of negligible mass and find

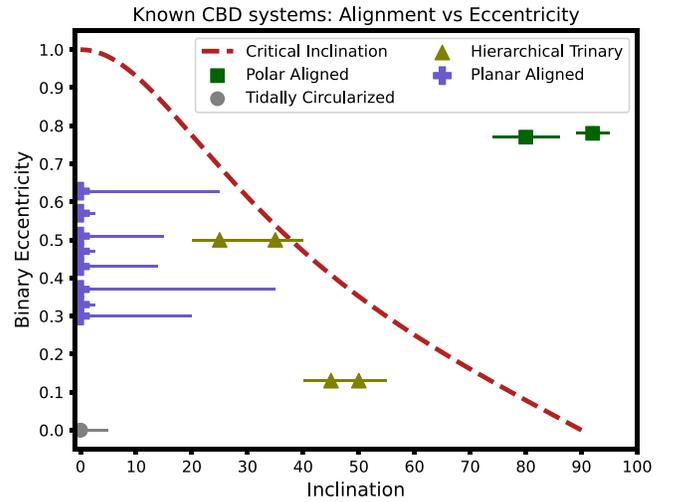

**Figure 1.** The red line gives the critical inclination angle as a function of eccentricity (Equation (1)). Observed systems from Table 1 that lie above and below the curve are predicted to evolve to polar and planar alignment, respectively. Green squares are systems that have approximately polar inclinations. Gray circles are short orbit systems expected to be circularized tidally. Olive triangles are hierarchical trinary systems. The presence of the third body influences the disk outside of simple predictions made for binary systems. Purple crosses are "aligned systems." They have no third body, and their inclination is below the critical curve. Excluding the triple systems, all other systems obey the predictions of R. G. Martin & S. H. Lubow (2019) that binaries will evolve either toward aligned or polar configurations, depending on which side of the critical curve they are initiated.

the following criterion:

$$i_{\min} = \arccos\left(\sqrt{\frac{5}{4e_b^2 + 1}}\, e_b\right). \qquad (1)$$

This provides a way to determine if any known CBDs are in polar alignment or will evolve to such a state at a later time. We plotted the eccentricity vs inclination angle for these known systems from Table 1 along with the curve we gained in Equation (1), and the result is shown in Figure 1. Discussion of these results will follow in Section 4.

For systems with zero inclination, we investigate how the eccentricity will evolve in time, using the predictions of M. Siwek et al. (2023) for the eccentricity evolution of aligned systems. Although these predictions rely on fixed binary 2D simulation results, we adopt them due to the subset of aligned systems falling below the critical inclination curve seen in Figure 1. Studies that allow live binaries and 3D effects find that for binaries with small inclination, $i \lesssim 20°$, eccentricity effects due to Kozai–Lidov oscillations are small. (R. G. Martin & S. H. Lubow 2019). Using the first table in Figure 3 from M. Siwek et al. (2023), we obtained values for the time derivative of eccentricity as a function of both eccentricity and mass ratio $q$. We interpolate the points where $\dot{e}_b = 0$ and plot them as an attractor curve. We interpolate points for $e_b = 0.7$ since those were not listed in the table. P. C. Duffell et al. (2020, Equation (15)) provides a form for the time derivative of mass ratio as a function of mass ratio:

$$\dot{q} = \frac{\dot{M}_b}{M_b}\frac{(1+q)(\lambda(q) - q)}{1 + \lambda(q)}, \quad \lambda(q) = \frac{1}{0.1 + 0.9q}. \qquad (2)$$

The values for $\dot{e}_b$ and the function for $\dot{q}$ allow for a vector map that gives the proposed evolution of a system's eccentricity and mass ratio further in time, as seen in

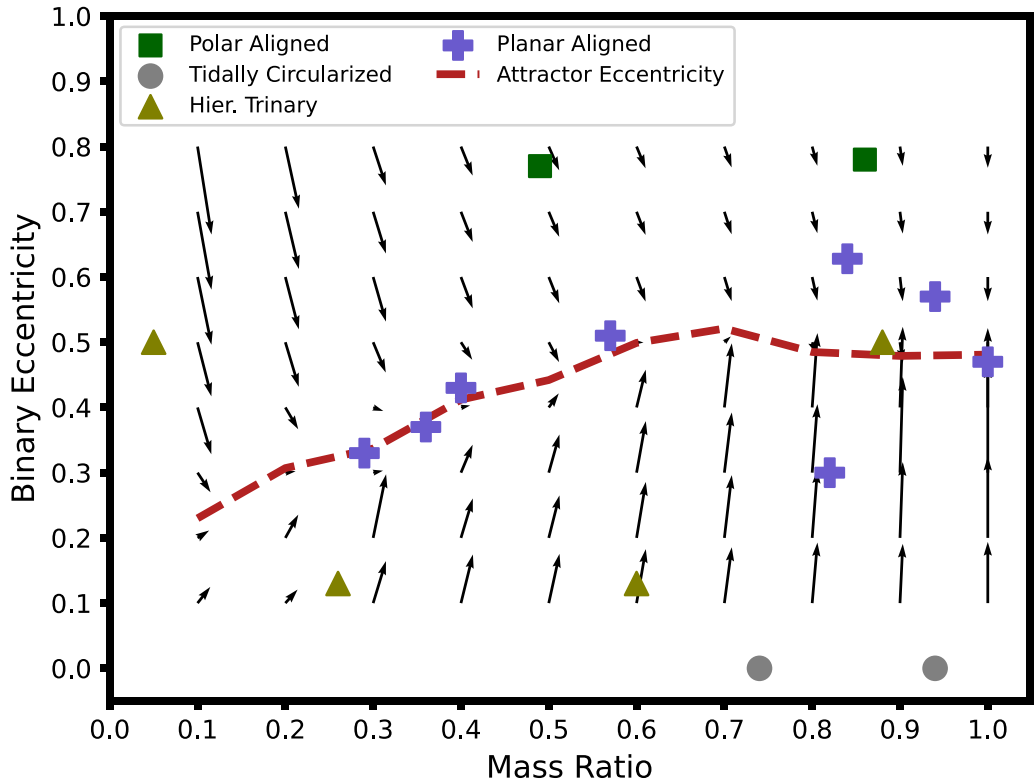

**Figure 2.** Vector map of the evolution of eccentricity and mass ratio in time as predicted by current aligned CBD theory. Observed systems are plotted to compare the aligned systems' distribution across $e$–$q$ space. Attractor eccentricity is plotted as a red line, and the observed systems are plotted as in Figure 1. The aligned systems from Table 1 lie along the curve as predicted by binary–disk interaction theory, assuming zero inclination.

Figure 2. For long-term evolution of binary–disk systems considering semimajor axis, mass ratio, and eccentricity of the binary, see R. Valli et al. (2024). R. Valli et al. (2024) look at simulation results from three separate studies (D. J. D'Orazio & P. C. Duffell 2021; J. Zrake et al. 2021; M. Siwek et al. 2023) and test consistency in CBD dynamics between the different parameter spaces used. They found that the different models provide consistent predictions with slight variations in behavior around $e \simeq 0.1$.

## 4. Results

Figure 1 has the critical inclination curve plotted with the known eccentricities and mutual inclination angles from Table 1. The systems are split into four separate categories based on the procedure in Section 2. Tidally circularized systems are plotted with gray circles. The systems that are polar aligned are plotted with green squares. Systems with a third body are plotted with olive triangles. The effect of the third body is likely significant enough to influence the orbit of the binary beyond the simple predictions from disk–binary theory. These systems have inclination evolution similar to CBD, but the timescale for evolution could be longer than the disk lifetime (S. Ceppi et al. 2023; S. Lepp et al. 2023). It is likely that this is the reason these are the only systems with modest inclinations. The remaining aligned systems are plotted with purple crosses.

Figure 1 is a confirmation of the predictions of R. G. Martin & S. H. Lubow (2019); systems to the left of the critical curve become aligned and systems to the right of the curve evolve into a polar configuration. The only exceptions are hierarchical trinaries, which are complex systems that can undergo apsidal precession of the disk and exhibit disk breaking. For a discussion of these effects, see S. Lepp et al. (2023) and S. Ceppi et al. (2023). These systems do not obey the same rules as predicted by calculations of a single binary interacting with a single disk.

Theoretical modeling (D. J. D'Orazio & P. C. Duffell 2021; J. Zrake et al. 2021; M. Siwek et al. 2023) has predicted that the aligned systems will migrate toward the attractor curve through time and once there, will move toward higher mass ratio along this curve until dissipation of the disk. Figure 2 shows that the eight aligned systems lie approximately upon the attractor eccentricity curve. These systems appear to have accreted enough gas to significantly affect their orbit. For example, system Ak Sco (S. H. P. Alencar et al. 2003; F. Anthonioz et al. 2015; I. Czekala et al. 2015), which has orbital parameters $q = 1.0$, $e = 0.47$ and an age of 18 Myr, is consistent with a twin binary that has experienced a great deal of accretion.

We find that observed aligned binary systems with no third body and in the absence of tidal circularization appear to fit the theoretically predicted curve of M. Siwek et al. (2023). Although this subset of systems lies in a steady state in regard to some of their parameters, this is not a useful extension to the entire population. See Figure 7, left from A. M. Price-Whelan et al. (2020) also Figure 11 in K. El-Badry et al. (2018); aside from the tidally circularized systems with $e_b = 0$, there is a fairly even distribution of eccentricities. There is likely some mechanism that drives eccentricity evolution after the disappearance of the disk.

## 5. Conclusions and Discussion

The abundance of stellar binaries makes them an adequate testing ground for binary–disk theory made to predict orbital evolution of SMBHBs. Although an individual SMBHB source has not yet been confirmed, studies such as M. Siwek et al. (2024) make predictions for gravitational-wave background and orbital parameters of SMBHBs with a CBD. By testing binary–disk interaction theory on existing stellar binary populations, we can help to bolster these arguments.

We analyzed current binary systems with CBDs to determine whether the disk is aligned with the orbital plane of the binary. We split the data set into four categories, depending on system properties and alignment. Some systems have perpendicularly aligned disks, some are tidally circularized, some are in hierarchical trinaries, and the rest were consistent with zero inclination. This is consistent with the theory of R. G. Martin & S. H. Lubow (2019). Binary systems that are above the critical inclination angle are found to be in polar alignment, whereas below this angle, they are all consistent with being aligned systems. The only exceptions are hierarchical trinaries whose outer binary–disk system can undergo the same inclination excitation, but the timescale for polar or planar alignment could be longer than the lifetime of the disk (S. Ceppi et al. 2023; S. Lepp et al. 2023).

Focusing on aligned systems, we find that their eccentricity distribution precisely obeys the predictions of D. J. D'Orazio & P. C. Duffell (2021), J. Zrake et al. (2021), M. Siwek et al. (2023), and R. Valli et al. (2024). Figure 2 highlights these systems with purple crosses. The vector map shows the evolution of the systems forward in time. We found that these systems all have modest eccentricities that lie close to the theoretically predicted attractor eccentricity of M. Siwek et al. (2023). However, current binary population in the Milky Way shows no preferential eccentricity for general systems. This suggests that there may be some other mechanism for eccentricity growth or randomization after dissipation of the disk.

One way to test this hypothesis would be to perform informed cuts on GAIA binary data, excluding triples and choosing a variety of cuts on stellar age, to see whether very young systems without a third body exhibit this same attractor eccentricity.

## Acknowledgments

We acknowledge R. Martin, M. Siwek, J. Zrake, and D. D'Orazio for their insightful discussion and suggestions. This research was funded by the National Science Foundation under grant number AAG-2206299. The analysis relied on the use of Python packages NumPy (C. R. Harris et al. 2020), Matplotlib (J. D. Hunter 2007), and Pandas (The pandas development team 2020). Finally, we would like to thank the anonymous referee for the careful review of our manuscript.

## ORCID iDs

Allen R. Murray  https://orcid.org/0009-0005-3325-4888
Paul C. Duffell  https://orcid.org/0000-0001-7626-9629